\documentclass[12pt]{article}
\input{epsf}
\newcommand{\vecsig}{\vec\sigma}
\newcommand{\veck}{\vec k}
\newcommand{\pwave}{$p$-wave}
\newcommand{\swave}{$s$-wave}
\newcommand{\Bmass}{M}
\newcommand{\Mmass}{m}
\newcommand{\Rscal}{\mu}

\begin{document}
\begin{center}
{\Large{\bf   Chiral dynamics of $p$-wave in $K^- p$ and coupled states
  }}

\vspace{0.3cm}

\end{center}

\vspace{1cm}

\begin{center}
{\large{D. Jido$^{a)}$\footnote{
email address: jido@rcnp.osaka-u.ac.jp, 
present address: 
        Research Center for Nuclear Physics (RCNP), 
        Ibaraki, Osaka 567-0047, Japan}, 
E. Oset$^{b)}$ and A. Ramos$^{a)}$}}
\end{center}

\begin{center}
{\small{$^{a)}$ \it 
     Departament d'Estructura i Constituents de la Mat\`{e}ria,\\ 
     Universitat de Barcelona, 
     Diagonal 647, 08028 Barcelona, Spain
}}

{\small{$^{b)}$ \it 
     Departamento de F\'{\i}sica Te\'orica and IFIC, \\
     Centro Mixto Universidad de Valencia-CSIC, \\
     Ap. Correos 22085, E-46071 Valencia, Spain }}

\end{center}

\vspace{1cm}

\begin{abstract}
    We perform an evaluation of the $p$-wave amplitudes of 
    meson-baryon scattering in the 
    strangeness $S=-1$ sector starting from the lowest order chiral 
    Lagrangians and introducing explicitly the $\Sigma^{*}$ field 
    with couplings to the meson-baryon states obtained using SU(6) 
    symmetry. The $N/D$ method of unitarization is used, 
    equivalent, in practice, to the use of the Bethe-Salpeter equation
    with a cut-off. The procedure leaves no freedom for the $p$-waves
    once the $s$-waves are fixed and thus one obtains genuine
    predictions for the $p$-wave scattering amplitudes, which are in
    good agreement with experimental results for differential cross
    sections, as well as for the width and partial decay widths of the
    $\Sigma^{*}$(1385). 
\end{abstract}

\section{Introduction}
The advent of chiral perturbation theory ($\chi PT$) as an effective
approach to QCD at low energies \cite{gasser} in hadron dynamics has
allowed steady progress in the field of meson-baryon interaction
\cite{ulf,veronique,pich,ecker}. Yet, an important step in
the application of the chiral Lagrangians at higher energies
than allowed by $\chi PT$ is the implementation of unitarity in
coupled channels. Pioneering works in this direction were those of
\cite{siegel,wolfram,kaiser}, where the Lippmann-Schwinger equation in
coupled channels was used extracting the kernels from the chiral
Lagrangians. Subsequent steps in this direction were done in
\cite{angels} in the study of the $K^- p$ interaction with the coupled
states using the Bethe-Salpeter equation and introducing all the
channels which could be formed from the octet of pseudoscalar mesons
and stable baryons. Further steps in this direction in the strangeness 
$S=0$ sector have been done in
\cite{om00,nicolajuan,juan,hosaka,inoue}. The works of
\cite{siegel,wolfram,kaiser,angels} deal only about the $s$-wave $K^-
p$ scattering, and one obtains a remarkably good agreement at low
energies with the data for transitions of $K^- p$ to different
channels, indicating that the $p$-wave and higher partial waves play a
minor role at these energies. The extension of these works to
include $p$-wave or higher partial waves is thus desirable in order to
see whether the agreement found with only $s$-wave is an accident or
whether one confirms that the contribution of the $p$-wave is indeed
small. There is also an important feature of the $p$-wave which is the
presence of the $\Sigma(1385)$ resonance appearing with the same
quantum numbers as those of the $K^- p$ system, although only visible
in $\pi\Lambda$ or $\pi\Sigma$ states since the resonance is below the
$K^- p$ threshold.

The introduction of $p$-waves in the strangeness $S=-1$ sector was done
in \cite{CaroRamon:2000jf} and more recently in
\cite{joseulf,Lutz:2002yb}. In \cite{CaroRamon:2000jf,joseulf} only
the region of energies above the $K^- p$ threshold is investigated but
the $\Sigma(1385)$ resonance region is not explored. In
\cite{Lutz:2002yb} the decouplet of the $\Sigma(1385)$ is explicitly
included as a field and new chiral Lagrangians to next to leading
order are introduced to deal with the meson-baryon scattering
problem. In this latter work the around 25 free parameters of the
theory are fitted to the data, although some of the parameters are
constrained by large $N_c$ arguments.

Simplicity is one of the appealing features of the $K^- p$
interactions from the perspective of chiral symmetry. Indeed, 
in \cite{angels}, it was found that using the transition amplitude
obtained with the lowest order chiral Lagrangian as a kernel of the
Bethe-Salpeter equation, and a cut-off of about 630 MeV to
regularize the loops, one could reproduce the cross sections of $K^- p
\rightarrow K^- p$, $\bar{K}^0n$, $\pi^0 \Lambda$, $\pi^0 \Sigma^0$,
$\pi^+ \Sigma^-$, $\pi^- \Sigma^+$, together with the properties of
the $\Lambda(1405)$ resonance, which is dynamically generated in that
scheme.

It is remarkable that, using the same input, one can also obtain
the $\Lambda(1670)$ and $\Sigma(1620)$ $s$-wave resonances
\cite{cornelius} as well as the $\Xi(1620)$ \cite{angelsnow}, which
completes the octet of lowest energy $s$-wave excited baryons together
with the $N^*(1535)$ obtained in \cite{siegel,juan,inoue} following
the same lines. 

The idea here is to see whether the simplicity observed in the
$s$-wave interaction holds also for $p$-waves. In other words, we would
like to see what does one obtain for the $p$-wave amplitudes using again
the lowest order chiral Lagrangians and the same cut-off parameter as
in \cite{angels}.  Anticipating results, we can say that the $p$-wave
amplitudes obtained with this line are in good agreement with
experiments, as well as the properties of the $\Sigma(1385)$
resonance, thus obtaining a parameter-free description of the $p$-wave
phenomenology in the $S=-1$ sector.

\section{Meson-baryon amplitudes to lowest order}
Following \cite{ulf,veronique,pich,ecker} we write the lowest order
chiral Lagrangian, coupling the octet of pseudoscalar mesons to the
octet of $1/2^+$ baryons, as

\begin{eqnarray}
   {\cal L}_1^{(B)} &=& \langle \bar{B} i \gamma^{\mu} \nabla_{\mu} B
    \rangle  - M \langle \bar{B} B\rangle  \nonumber \\
    &&  + \frac{1}{2} D \left\langle \bar{B} \gamma^{\mu} \gamma_5 \left\{
     u_{\mu}, B \right\} \right\rangle + \frac{1}{2} F \left\langle \bar{B}
     \gamma^{\mu} \gamma_5 \left[u_{\mu}, B\right] \right\rangle  
    \label{chiralLag}
\end{eqnarray}
where the symbol $\langle \, \rangle$ denotes the trace of SU(3) flavor
matrices, $M$ is the baryon mass and
\begin{eqnarray}
  \nabla_{\mu} B &=& \partial_{\mu} B + [\Gamma_{\mu}, B] \nonumber \ ,\\
  \Gamma_{\mu} &=& \frac{1}{2} (u^\dagger \partial_{\mu} u + u\, \partial_{\mu}
      u^\dagger) \nonumber \ , \\
  U &=& u^2 = {\rm exp} (i \sqrt{2} \Phi / f) \ , \\
  u_{\mu} &=& i u ^\dagger \partial_{\mu} U u^\dagger \nonumber \ . 
\end{eqnarray}
The couplings $D$ and $F$ are chosen as $D=0.85$, $F=0.52$.

The meson and baryon fields in the SU(3) matrix form are given by
\begin{equation}
\Phi =
\left(
\begin{array}{ccc}
\frac{1}{\sqrt{2}} \pi^0 + \frac{1}{\sqrt{6}} \eta & \pi^+ & K^+ \\
\pi^- & - \frac{1}{\sqrt{2}} \pi^0 + \frac{1}{\sqrt{6}} \eta & K^0 \\
K^- & \bar{K}^0 & - \frac{2}{\sqrt{6}} \eta
\end{array}
\right) \ ,
\end{equation}
\begin{equation}
B =
\left(
\begin{array}{ccc}
\frac{1}{\sqrt{2}} \Sigma^0 + \frac{1}{\sqrt{6}} \Lambda &
\Sigma^+ & p \\
\Sigma^- & - \frac{1}{\sqrt{2}} \Sigma^0 + \frac{1}{\sqrt{6}} \Lambda & n \\
\Xi^- & \Xi^0 & - \frac{2}{\sqrt{6}} \Lambda
\end{array}
\right) \ .
\end{equation}

The $BB\Phi\Phi$ interaction Lagrangian comes from the $\Gamma_{\mu}$ term
in the covariant derivative and we find
\begin{equation}
   {\cal L}_1^{(B)} = \left\langle \bar{B} i \gamma^{\mu} \frac{1}{4 f^2}
   [(\Phi\, \partial_{\mu} \Phi - \partial_{\mu} \Phi \Phi) B
   - B (\Phi\, \partial_{\mu} \Phi - \partial_{\mu} \Phi \Phi)] 
   \right\rangle \ , \label{lowest}
\end{equation}
from where one derives the transition amplitudes
\begin{equation}
   V_{ij} = - C_{ij} {1 \over 4 f^2} \bar{u}(p^\prime) \gamma^\mu u(p)
   (k_\mu + k^\prime_\mu)  \label{fourpoint}
\end{equation}
where $k$, $k^\prime$ ($p$, $p^\prime$) are the initial and final
meson (baryon) momenta, respectively, and the coefficients $C_{ij}$, 
where $i$, $j$ indicate the particular meson-baryon channel, form a
symmetric matrix and are written explicitly in
\cite{angels}. Following \cite{angels}, the meson decay constant $f$
is taken as an average value $f=1.123 f_\pi$ \cite{cornelius}. The
channels included in our study are $K^- p$, $\bar{K}^0n$, $\pi^0
\Lambda$, $\pi^0 \Sigma^0$, $\eta \Lambda$, $\eta \Sigma^0$, $\pi^+
\Sigma^-$, $\pi^- \Sigma^+$, $K^+ \Xi^-$, $K^0 \Xi^0$.  The $s$-wave
amplitudes are obtained in \cite{angels,cornelius} and we do not
repeat them here. The Lagrangian of eq.\ (\ref{lowest}) provides also
a small part of the $p$-wave which is easily obtained by evaluating
the matrix elements of eq.\ (\ref{fourpoint}) using the
explicit form of the spinor and the Dirac matrices. We obtain, in the
center of mass system, 
\begin{equation}
   t^{\, c}_{ij} = - C_{ij} {1 \over 4 f^2}\, a_i\, a_j \left({1 \over
   b_i} + {1 \over b_j} \right) (\vec\sigma \cdot \vec k_j)
   (\vec\sigma \cdot \vec k_i) \label{pwcont}
\end{equation}
with 
\begin{equation}
   a_i = \sqrt{E_i + M_i \over 2 M_i}\ , \hspace{0.7cm} b_i = E_i +
   M_i\ , \hspace{0.7cm} E_i = \sqrt{M_i^{\, 2} + \vec p_i^{\, 2}}
\end{equation}
where $M_i$ is the mass of the baryon in channel $i$.

\begin{figure}
   \begin{center}
   \epsfxsize=13cm
   \epsfbox{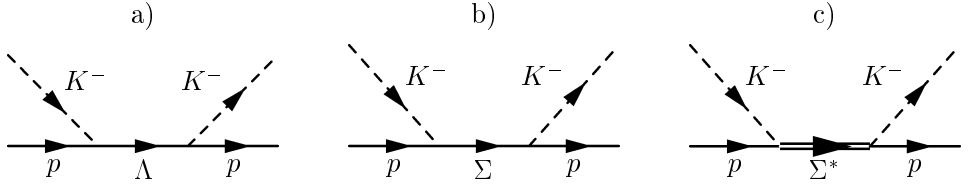}
   \end{center}
   \caption{Diagrams for the pole terms of a) $\Lambda$, b) $\Sigma$
   and c) $\Sigma^*$ with the $K^- p$ channel as an
   example. \label{fig:poleterm}}  
\end{figure}

In addition we have the contribution from the $\Lambda$ and $\Sigma$
pole terms which are obtained from the $D$ and $F$ terms of the
Lagrangian of eq.~(\ref{chiralLag}). The $\Sigma^*$ pole term is also
included explicitly with couplings to the meson-baryon states
evaluated using SU(6) symmetry arguments \cite{Oset:2001eg}.
These terms correspond to the diagrams of Fig.~\ref{fig:poleterm} a),
b), c). 

The amplitudes for the $\Lambda$, $\Sigma$ and $\Sigma^*$ pole terms
are readily evaluated and performing a nonrelativistic reduction,
keeping terms up to ${\cal O}(p/M)$, we find in the center of mass
frame,
\begin{eqnarray}
   t^{\Lambda}_{ij} &=& D^{\Lambda}_i D^{\Lambda}_j { 1 \over \sqrt{s}
   - \tilde M_\Lambda} (\vecsig \cdot \veck_j)(\vecsig \cdot \veck_i)
   \left(1+{k_j^0 \over M_j}\right) \left(1+{k_i^0 \over M_i} \right) 
   \nonumber \ , \\
   t^{\Sigma}_{ij} &=& D^{\Sigma}_i D^{\Sigma}_j { 1 \over \sqrt{s}
   - \tilde M_\Sigma} (\vecsig \cdot \veck_j)(\vecsig \cdot \veck_i)
   \left(1+{k_j^0 \over M_j} \right) \left(1+{k_i^0 \over M_i} \right) 
   \label{poleamps}\ , \\
   t^{\Sigma^*}_{ij} &=& D^{\Sigma^*}_i D^{\Sigma^*}_j { 1 \over \sqrt{s}
   - \tilde M_{\Sigma^*}} (\vec S \cdot \veck_j)(\vec S^\dagger \cdot
   \veck_i) \nonumber  
\end{eqnarray}
with $S^\dagger$ the spin transition operator form spin $1/2$ to $3/2$
with the property
\begin{equation}
    \sum_{M_{s}} S_{i} | M_{s} \rangle \langle M_{s} | 
    S^{\dagger}_{j} = {2 \over 3} \delta_{ij} - { i \over 3 } 
    \epsilon_{ijk} \sigma_{k}
\end{equation}
and 
\begin{eqnarray}
   D^\Lambda_i &=& c_i^{D,\Lambda} \sqrt{20 \over 3} {D \over 2 f} -
   c_i^{F,\Lambda} \sqrt{12} { F \over 2 f} \nonumber \ , \\
   D^\Sigma_i &=& c_i^{D,\Sigma} \sqrt{20 \over 3} {D \over 2 f} -
   c_i^{F,\Sigma} \sqrt{12} { F \over 2 f} \ , \\
   D^{\Sigma^*}_i &=& c_i^{S,\Sigma^*} {12 \over 5} {D + F\over 2 f}
   \nonumber \ .
\end{eqnarray}
The constants $c^D$, $c^F$, $c^S$ are SU(3) Clebsch-Gordan
coefficients which depend upon the meson and baryon involved in the
vertex and are given in Table~\ref{tab:coffs}. The phase relating
physical states to isospin states $|K^- \rangle = - | 1/2,1/2 \rangle$,
$ |\Xi^- \rangle = - |1/2,1/2 \rangle$, $ |\pi^+ \rangle = -|1,1
\rangle$, $ |\Sigma^+ \rangle = -|1,1 \rangle$, normally adopted in
the chiral Lagrangians, are also assumed here. $\tilde M_\Lambda$,
$\tilde M_\Sigma$, $\tilde M_{\Sigma^*}$ are bare masses of the
hyperons, which will turn into physical masses upon unitarization.

\begin{table}
  \begin{center}
  \begin{tabular}{c|rrrrrrrrrr} 
        & $K^-p$ & $\bar K^0 n$ & $\pi^0 \Lambda$ & $\pi^0 \Sigma^0$ &
        $\eta \Lambda$ & $\eta \Sigma^0$ &
         $\pi^+ \Sigma^-$ & $\pi^- \Sigma^+$ & $K^+\Xi^-$ & $K^0\Xi^0$ \\
     \hline
     $c_i^{D,\Lambda}$ & $-\sqrt{1 \over 20}$ & $-\sqrt{1 \over 20}$ &
         $0$ & $\sqrt{1 \over 5}$ & $-\sqrt{1 \over 5}$ & $0$ & 
         $\sqrt{1 \over 5}$ & $\sqrt{1 \over 5}$ & 
         $-\sqrt{1 \over 20}$ & $-\sqrt{1 \over 20}$ \\
     $c_i^{F,\Lambda}$& $\sqrt{1 \over 4}$  & $\sqrt{1 \over 4}$ & 
         $0$ & $0$ & $0$ & $0$ & $0$ & $0$ & 
         $-\sqrt{1 \over 4}$ & $-\sqrt{1 \over 4}$ \\
     $c_i^{D,\Sigma}$ & $\sqrt{3 \over 20}$ & $-\sqrt{3 \over 20}$ & 
         $\sqrt{1 \over 5}$ & $0$ & $0$ & $\sqrt{1 \over 5}$ & $0$ & $0$ &
         $\sqrt{3 \over 20}$ & $-\sqrt{3 \over 20}$ \\
     $c_i^{F,\Sigma}$ & $\sqrt{1 \over 12}$ & $-\sqrt{1 \over 12}$ &
         $0$ & $0$ & $0$ & $0$ & $-\sqrt{1 \over 3}$ & $\sqrt{1 \over 3}$ & 
        $-\sqrt{1 \over 12}$ & $\sqrt{1 \over 12}$ \\
     $c_i^{S,\Sigma^*}$ & $-\sqrt{1 \over 12}$ & $\sqrt{1 \over 12}$ & 
        $\sqrt{1 \over 4}$ & $0$ & $0$ & $-\sqrt{1 \over 4}$ & 
        $-\sqrt{1 \over 12}$ & $\sqrt{1 \over 12}$ &
        $\sqrt{1 \over 12}$ & $-\sqrt{1 \over 12}$
  \end{tabular}
  \caption{$c^D$, $c^F$, $c^S$ coefficients of eq.\ (\ref{poleamps}).
  \label{tab:coffs}}
  \end{center}
\end{table}

\section{Unitary amplitudes}

The lowest order (tree level) contributions to the $p$-wave
meson-baryon scattering matrix are provided by eqs.\ (\ref{pwcont}) and
(\ref{poleamps}).  Following the philosophy of \cite{angels}, the tree
level contributions are used as a kernel of the Bethe-Salpeter
equation. 
Furthermore, the factorization of the kernel makes it technically simpler
to solve the Bethe-Salpeter equation. It was shown in 
\cite{angels} that the kernel for the $s$-wave amplitudes can
be factorized on the mass shell in the loop functions, by making some
approximations typical of heavy baryon perturbation theory.  
The factorization for {\pwave}s in
meson-meson scattering is also proved in \cite{daniel} along the same
lines. A different, more general, proof of the factorization is done
in \cite{joseantonio} for meson-meson interactions and in
\cite{joseulf} for meson-baryon ones where, using the $N/D$ method of
unitarization and performing dispersion relations, one proves the
on-shell factorization in the absence of the left-hand cut
contribution. This part is shown to be small in  
\cite{joseantonio} for meson-meson scattering and even smaller for the
meson-baryon case because of the large baryonic masses in the
meson-baryon systems. The formal result obtained in \cite{joseulf} for
the meson-baryon amplitudes in the different channels is given in
matrix form by the same result coming from the Bethe-Salpeter equation 
\begin{equation}
    T = V + V G T \ ,
\end{equation}
that is 
\begin{equation}
    T = [1-VG]^{-1} V \label{BSeq}
\end{equation}
where $V$ is the kernel (potential), given by the amplitudes of eqs.\ 
(\ref{pwcont}) and (\ref{poleamps}), and $G$ is a diagonal matrix 
accounting for the loop function of a meson-baryon propagator, which 
must be regularized to eliminate the infinities.  In \cite{angels} it 
was regularized by means of a cut-off, while in \cite{joseulf} 
dimensional regularization was used. Both methods are eventually 
equivalent although in the dimensional regularization scheme there is 
a different subtraction constant in each isospin channel and thus 
allows for more freedom. 

The loop function $G$ in the cut-off method is given by
\begin{eqnarray}
    G_{l}(\sqrt{s}) &=& i \int {d^{4} q \over (2 \pi)^{4}} { M_{l} \over 
    E_{l}(\vec q) } { 1 \over k^{0}+p^{0}-q^{0}-E_{l}(\vec q) + i 
    \epsilon} { 1 \over q^{2} - m_{l}^{2} + i\epsilon} \\
    &=& \int^{q_{\rm max}} {d^{3} q \over (2 \pi)^{3}} { 1 \over 2 
    \omega_{l}(q)} {M_{l} \over E_{l}(q)} { 1 \over 
    p^{0}+k^{0}-\omega_{l}(q)-E_{l}(q) } \nonumber
\end{eqnarray}
while in dimensional regularization one has
\begin{eqnarray}
    G_l(\sqrt{s}) 
    &=& i 2 \Bmass_l \int {d^4 q \over (2\pi)^4 } { 1 \over
    (P-q)^2 - \Bmass_l^2 + i \epsilon}{1 \over q^2 - \Mmass_l^2 +
    i\epsilon} \nonumber \\
    &=& {2 \Bmass_l \over 16 \pi^2} \left\{ a(\Rscal) + \ln {\Bmass^2_l
    \over \Rscal^2} + {\Mmass_l^2 - \Bmass_l^2 + s \over 2 s} \ln
    {\Mmass^2_l \over \Bmass^2_l} \right. \label{gfunc} \\  
    && + {\bar q_l \over \sqrt{s} } \left[ \ln(s - (\Bmass^2_l -
    \Mmass^2_l) + 2 \bar q_l \sqrt{s}) + \ln(s + (\Bmass^2_l -
    \Mmass^2_l) + 2 \bar q_l \sqrt{s}) \right. \nonumber \\
    && \left. \left.  - \ln(- s + (\Bmass^2_l - \Mmass^2_l) + 2 \bar
    q_l \sqrt{s}) - \ln(- s - (\Bmass^2_l - \Mmass^2_l) + 2 \bar q_l
    \sqrt{s}) \right] \right\} \ , \nonumber
\end{eqnarray}
where $\Mmass$ and $\Bmass$ are taken to be the observed meson and baryon
masses, respectively, in order to respect the phase space allowed by
the physical states, and $\mu$ is a regularization scale (playing the 
role of a cut-off) and $a_{i}$ are subtraction constants in each of 
the isospin channels.  In \cite{cornelius} it was shown that, taking 
$\mu=630$ MeV as the cut-off in \cite{angels}, the values of the 
subtraction constants in eq.\ (\ref{gfunc}) which lead to the same 
results as in the cut-off scheme are \footnote{The $a_{K\Xi}$
   parameter quoted here is slightly changed from $-2.56$ in
   \cite{cornelius} in order to get the position of the
   $\Lambda$(1670) resonance better, but as show there, this
   parameters has no relevance in the low energy results studied
   here.} 
\begin{eqnarray}
   a_{\bar K N} = -1.84, \hspace{0.5cm} a_{\pi \Sigma} = - 2.00,
   \hspace{0.5cm} a_{\pi \Lambda} = -1.83 \nonumber \\
   a_{\eta \Lambda} = -2.25, \hspace{0.5cm} a_{\eta \Sigma} = -2.38,
   \hspace{0.5cm} a_{K \Xi} = -2.67\ .
   \label{subtsorig}
\end{eqnarray}
We shall use the same values here and hence this procedure would be 
equivalent to performing the calculations using  a unique cut-off 
$q_{\rm max}=630$ MeV in all channels. In a second step we shall relax 
this constraint and allow the subtraction constants to vary freely to 
obtain a better global fit to the data.

The use of the amplitudes of eqs.\ (\ref{pwcont}) and (\ref{poleamps})
directly in eq.\ (\ref{BSeq}) is impractical since, due to their spin 
structure, there is a mixture of different angular momenta.  
It is standard to separate the amplitude for a spin zero meson and a
spin 1/2 baryon into different angular momentum components. We write,
with the angle $\theta$ between meson momenta in initial and final 
states, 
\begin{eqnarray}
    f(\veck^{\prime}, \veck) & = &\sum_{l=0}^{\infty} \left\{ (l+1) 
    f_{l+} + l f_{l-} \right\} P_{l}(\cos \theta) \nonumber \\
    && -i \vecsig 
    \cdot ( \hat k^{\prime} \times \hat k) \sum_{l=0}^{\infty} \left\{ 
    f_{l+} - f_{l-} \right\} P^{\prime}_{l}(\cos \theta) \ ,
    \label{partwaveamp}
\end{eqnarray}
which separates the amplitude into a spin non-flip part and a spin
flip one.  The amplitudes $f_{l+}$, $f_{l-}$ correspond to orbital
angular momentum $l$ and total angular momentum $l+1/2$, $l-1/2$,
respectively.  This amplitudes exhibit independent unitarity
conditions and separate in the Bethe-Salpeter equation.  If we specify 
the $l=1$ case, the $p$-wave amplitudes can be written as
\begin{equation}
    T(\veck^\prime, \veck) = (2l +1) \left( f(\sqrt{s})\, \hat
    k^{\prime} \cdot \hat k - i g(\sqrt{s})\, (\hat k^{\prime} \times
    \hat k) \cdot \vecsig \right) \hspace{0.7cm} (l=1) \ . \label{pwamp}
\end{equation}
From Eqs.~(\ref{pwcont}), (\ref{poleamps}) the corresponding lowest order 
(tree level) amplitudes read 
\begin{eqnarray}
    f^{\rm tree}_{ij}(\sqrt{s}) &=& {1 \over 3} \left\{ - C_{ij} {1 \over 
4 f^2}\, a_i\,
    a_j \left({1 \over b_i} + {1 \over b_j} \right) 
    + { D^{\Lambda}_i D^{\Lambda}_j \left(1+{k_i^0 \over M_i} \right) 
    \left(1+{k_j^0 \over M_j} \right) \over \sqrt{s} - \tilde M_\Lambda} 
    \right.  \nonumber \\
    && \left.  + { D^{\Sigma}_i D^{\Sigma}_j \left(1+{k_i^0 \over M_i}
    \right) \left(1+{k_j^0 \over M_j} \right) \over \sqrt{s} - \tilde
    M_\Sigma} 
    + {2 \over 3} {D^{\Sigma^{*}}_i D^{\Sigma^{*}}_j \over
    \sqrt{s} - \tilde M_\Sigma^{*}} \right\} k_{i} k_{j}
    \\
    g^{\rm tree}_{ij}(\sqrt{s}) &=& {1 \over 3} \left\{  C_{ij} {1 \over 4 
f^2}\, a_i\,
    a_j \left({1 \over b_i} + {1 \over b_j} \right) 
    - { D^{\Lambda}_i D^{\Lambda}_j \left(1+{k_i^0 \over M_i} \right) 
    \left(1+{k_j^0 \over M_j} \right) \over \sqrt{s} - \tilde M_\Lambda} 
    \right.   \nonumber \\
    && \left.  - { D^{\Sigma}_i D^{\Sigma}_j \left(1+{k_i^0 \over M_i}
    \right) \left(1+{k_j^0 \over M_j} \right) \over \sqrt{s} - \tilde
    M_\Sigma} + {1 \over 3} {D^{\Sigma^{*}}_i D^{\Sigma^{*}}_j \over
    \sqrt{s} - \tilde M_\Sigma^{*}} \right\} k_{i} k_{j}
\end{eqnarray}
where $i,j$ are channel indices. Hence, denoting $f_{l+} \equiv 
f_{+}$ , $f_{l-} \equiv f_{-}$ for $l=1$, with
\begin{eqnarray}
    f_{+} &=& f+g \label{fg} \\
    f_{-} &=& f-2g \nonumber \ ,
\end{eqnarray}
and using Eq.~(\ref{BSeq}), one obtains
\begin{eqnarray}
    f_{+} &=& [1-f_{+}^{\rm tree} G ]^{-1} f_{+}^{\rm tree}  
    \label{pwBSeq} \\
    f_{-} &=& [1-f_{-}^{\rm tree} G ]^{-1} f_{-}^{\rm tree} \ . \nonumber
\end{eqnarray}
As one can see from these equations, 
the amplitudes $f_{+}^{\rm tree}$, $f_{-}^{\rm  tree}$ in the 
diagonal meson-baryon channels contain the factor $\veck^{2}$, with 
$\veck$ the 
on-shell 
center-of-mass momentum of the meson in this channel. For transition 
matrix elements from channel $i$ to $j$ the corresponding factor is 
$k_{i}k_{j}$ where the energy and momentum of the meson in a certain 
channel are given by 
\begin{equation}
    E_{i} = { s + m_{i}^{2} - M_{i}^{2} \over 2 \sqrt{s}} \ ; 
    \hspace{0.5cm} k_{i} = \sqrt{E_{i}^{2} - m_{i}^{2}} \ , 
\end{equation}
which also provides the analytical extrapolation below the threshold 
of the channel where $k_i$ becomes purely imaginary. 

The differential cross sections, including the \swave\ amplitudes are 
given by
\begin{equation}
    {d \sigma_{ij} \over d\Omega} = {1 \over 16 \pi^{2} } {M_{i} M_{j} 
    \over s} {k^{\prime} \over k} \left\{ \left| f^{(s)} + (2 f_{+} + 
    f_{-}) \cos \theta \right|^{2} + \left| f_{+} - f_{-} \right|^{2} 
    \sin^{2}\theta \right\}
\end{equation}
where the subscript $i$, $j$ in each of the amplitudes must be 
understood. The set of equations (\ref{pwBSeq}) can be solved in the space 
of physical states, the ten-channel space introduced in the former 
section. 
Alternatively one can also construct states of given isospin (see 
section 3 of Ref.~\cite{angels}) and work directly with isospin states. 
Conversely, one can work with the physical states and construct the 
isospin amplitudes from the appropriate linear combinations of 
transition amplitudes with physical states. The isospin separation 
is useful for \pwave. Indeed in the channel $f_{-}$, which corresponds 
to $J=1/2$, we can have $I=0$, $I=1$. The pole of the $\Lambda$ and 
the $\Sigma$ from the pole terms in Fig.~\ref{fig:poleterm} will show 
up in the calculation in these channels, respectively.  However, the 
unitarization procedure will shift the mass from a starting bare 
mass $\tilde M_{\Lambda}$, $\tilde M_{\Sigma}$ in eqs.\ 
(\ref{poleamps}) to another mass which we demand to be the physically 
observed mass.  Similarly, in the $f_{+}$ amplitude, corresponding 
to $J=3/2$, there will be a pole for $I=1$ corresponding to the 
$\Sigma^{*}$. Once again we start from a bare mass $\tilde 
M_{\Sigma^{*}}$ in eqs.\ (\ref{poleamps}) such that after unitarization 
the pole appears at the physical $\Sigma^{*}$ mass. In the case of 
the $\Sigma^{*}$, since there is phase space for decay into $\pi 
\Sigma$ and $\pi \Lambda$, the unitarization procedure will 
automatically provide the width of the $\Sigma^{*}$. With no free 
parameters to play with, the results obtained for the $\Sigma^{*}$ 
width and the branching ratios to the $\pi\Sigma$ and $\pi \Lambda$ 
channels will be genuine predictions of the theoretical framework. 
Since the poles of the coupled channel $T$ matrix appear when the 
determinant of the 
$[1-f_{+}^{\rm tree}G]$ or $[1-f_{-}^{\rm tree} G]$ matrices is zero 
(in the complex plane),
it is clear from Eq.~(\ref{pwBSeq}) that one gets the same $\Lambda$, 
$\Sigma$ or $\Sigma^{*}$ 
poles in all the matrix elements.
\section{Results}

\begin{figure}
   \begin{center}
   \epsfxsize=13cm
   \epsfbox{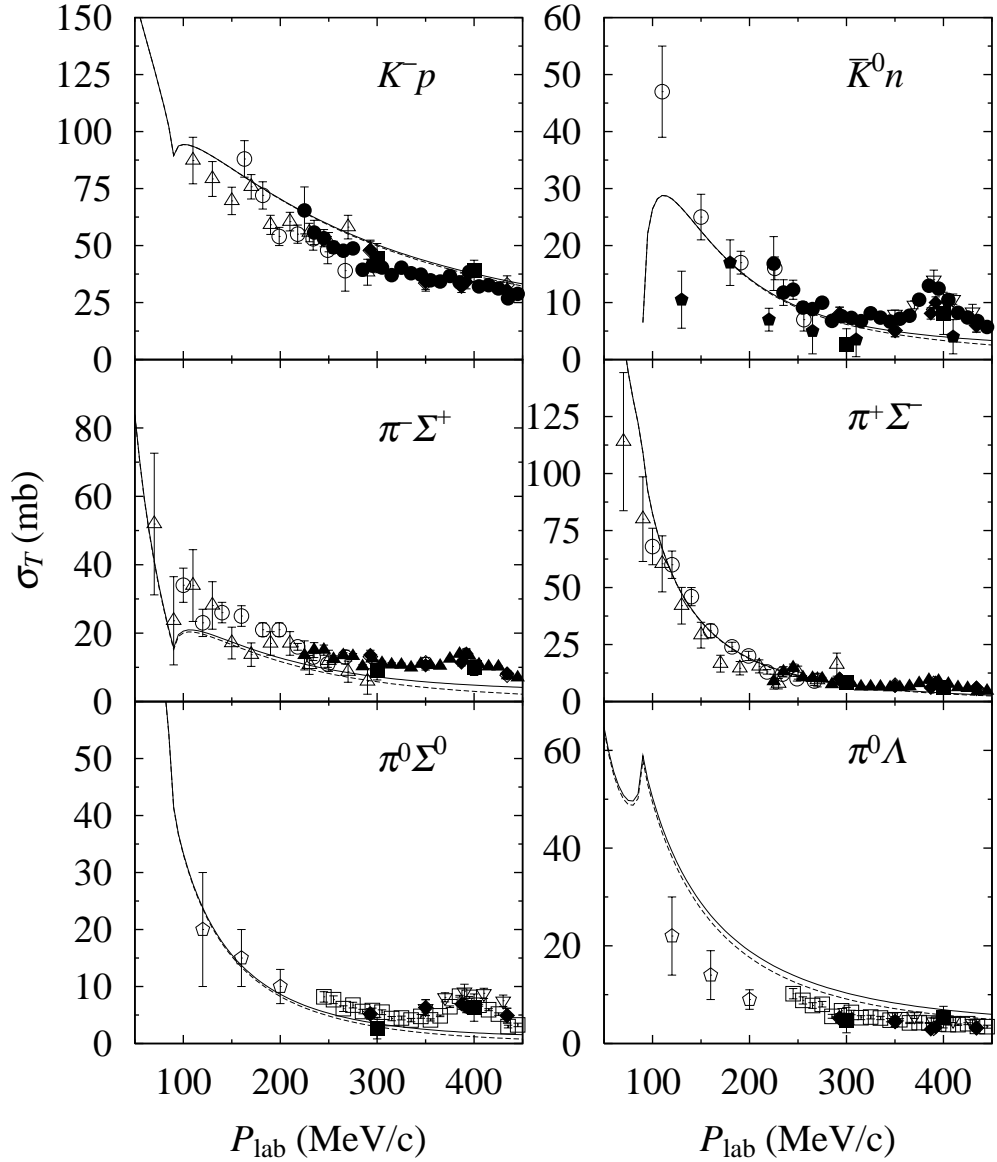}
   \end{center}
   \caption{Total cross sections of the $K^{-}p$ elastic and 
   inelastic scatterings. The solid line denotes our results with the
   parameter set of Eq.~(\ref{subtsorig}) including both \swave\ and 
\pwave.
   The dashed line shows our results without the \pwave\ amplitudes.
   The data are taken from : \cite{Ciborowski:1982et} (open circles),
   \cite{Bangerter:1981px} (black triangles), \cite{Mast:1976pv} (black
   circles), \cite{Sakitt:1965kh} (open triangles), \cite{Mast:1975sx}
   (open squares), \cite{Nordin61} (black squares),
   \cite{Berley:1970zh} (open down triangles), \cite{ferrolozzi62}
   (open diamonds), \cite{watson63} (black diamonds),
   \cite{eberhard59} (open pentagons) and \cite{kim65} (black
   pentagons). 
   \label{fig:totalorig}}  
\end{figure}

In Fig.~\ref{fig:totalorig}, we can see the total cross sections for 
$K^{-}p$ to different channels as a function of the initial meson 
momentum in the laboratory frame.  The parameters taken there are the
set of values of the subtraction constants $a_i$ from eq.\
(\ref{subtsorig}) which are already fixed from \cite{cornelius}, and
the values of the bare masses, $\tilde M_{\Lambda} = 1078$ MeV,
$\tilde M_{\Sigma} = 1104$ MeV, $\tilde M_{\Sigma^{*}}=1359$ MeV. The
results with the \swave\  alone are equivalent to those presented in
\cite{angels}. As already remarked there, the agreement with
experiment is quite good, particularly taking into account that only a
free parameter, the cut-off, has been fitted to the data. In addition
the threshold ratios $\gamma$, $R_{c}$, $R_{n}$ defined as 
\begin{eqnarray}
\gamma &=& \frac{\Gamma (K^- p \rightarrow \pi^+ \Sigma^-)}
{\Gamma (K^- p \rightarrow \pi^- \Sigma^+)}   = 2.36 \pm 0.04
\nonumber \\
R_c &=& \frac{\Gamma (K^- p \rightarrow \hbox{charged particles)}}
{\Gamma (K^- p \rightarrow \hbox{all)}}  = 0.664 \pm 0.011 
\\
R_n &=& \frac{\Gamma (K^- p \rightarrow \pi^0 \Lambda)}
{\Gamma (K^- p \rightarrow \hbox{all neutral states})} = 0.189 \pm 0.015
\nonumber
\end{eqnarray}
were also well reproduced. The values obtained here are $\gamma=2.30$, 
$R_{c}=0.618$, $R_{n}=0.257$. 

\begin{figure}
   \begin{center}
   \epsfxsize=13cm
   \epsfbox{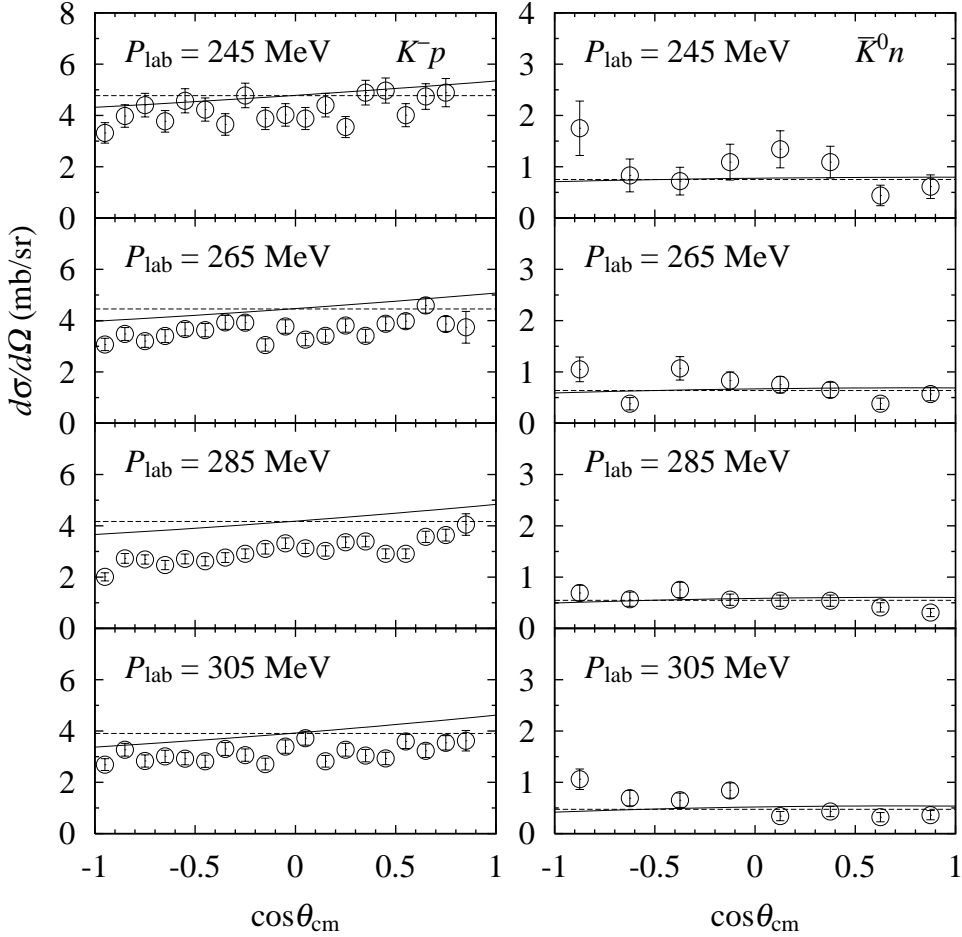}
   \end{center}
   \caption{Differential cross sections of the $K^{-}p \rightarrow K^-
   p$, $\bar{K}^0 n$ scatterings at $p_{\rm lab}=$ 245, 265, 285 and
   305 MeV. The solid line denotes our results with the parameter set
   of Eq.~(\ref{subtsorig}) including both \swave\ and \pwave.  The dashed 
   line shows our results without the \pwave\ amplitudes. The data are taken
   from \cite{Mast:1976pv}. \label{fig:difforig}}  
\end{figure}

\begin{figure}
   \begin{center}
   \epsfxsize=13cm
   \epsfbox{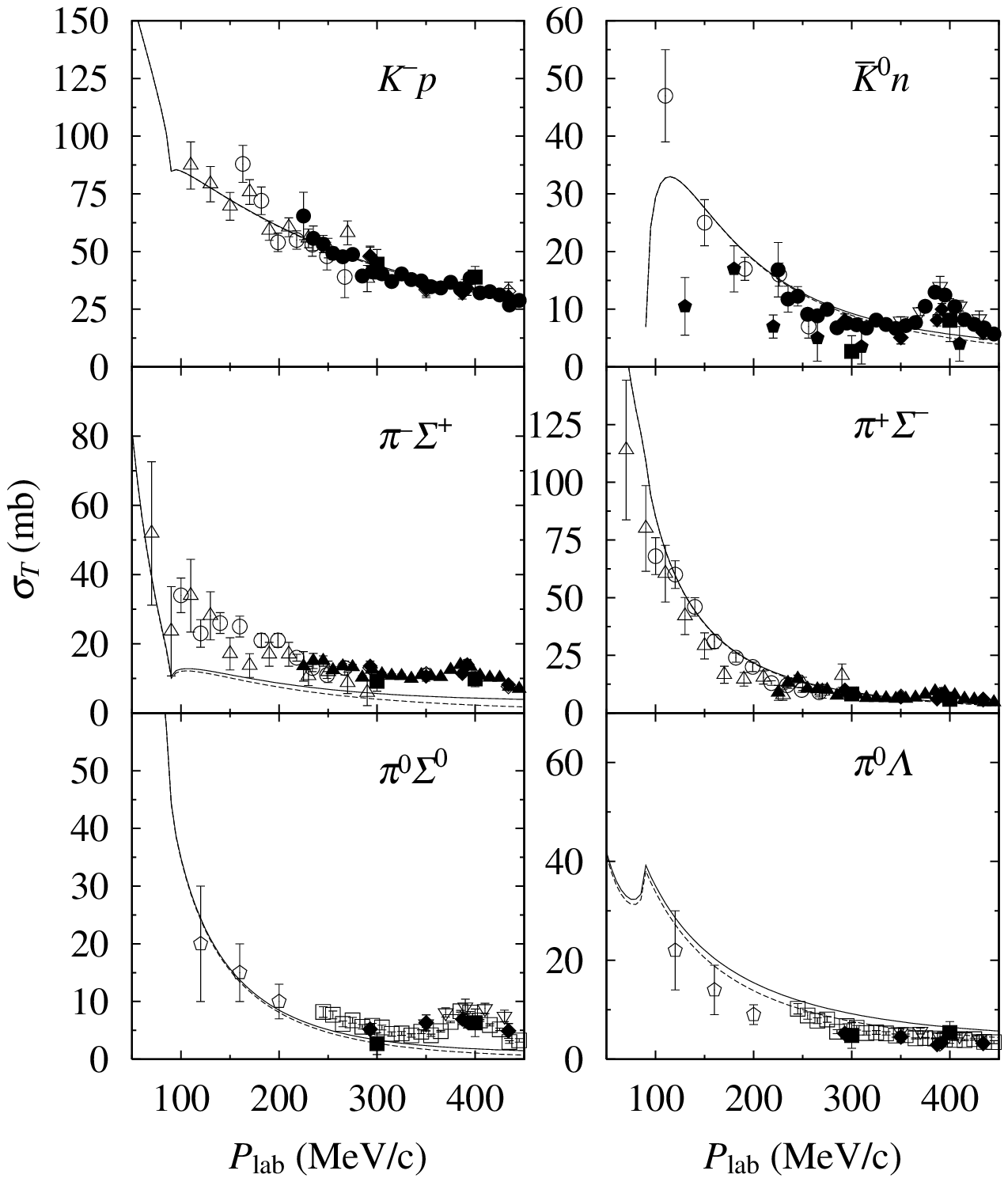}
   \end{center}
   \caption{
    Same as Fig.~\ref{fig:totalorig} with the parameter set of
   Eq.~(\ref{subtsnew}).  
   \label{fig:totalnew}}  
\end{figure}

\begin{figure}
   \begin{center}
   \epsfxsize=13cm
   \epsfbox{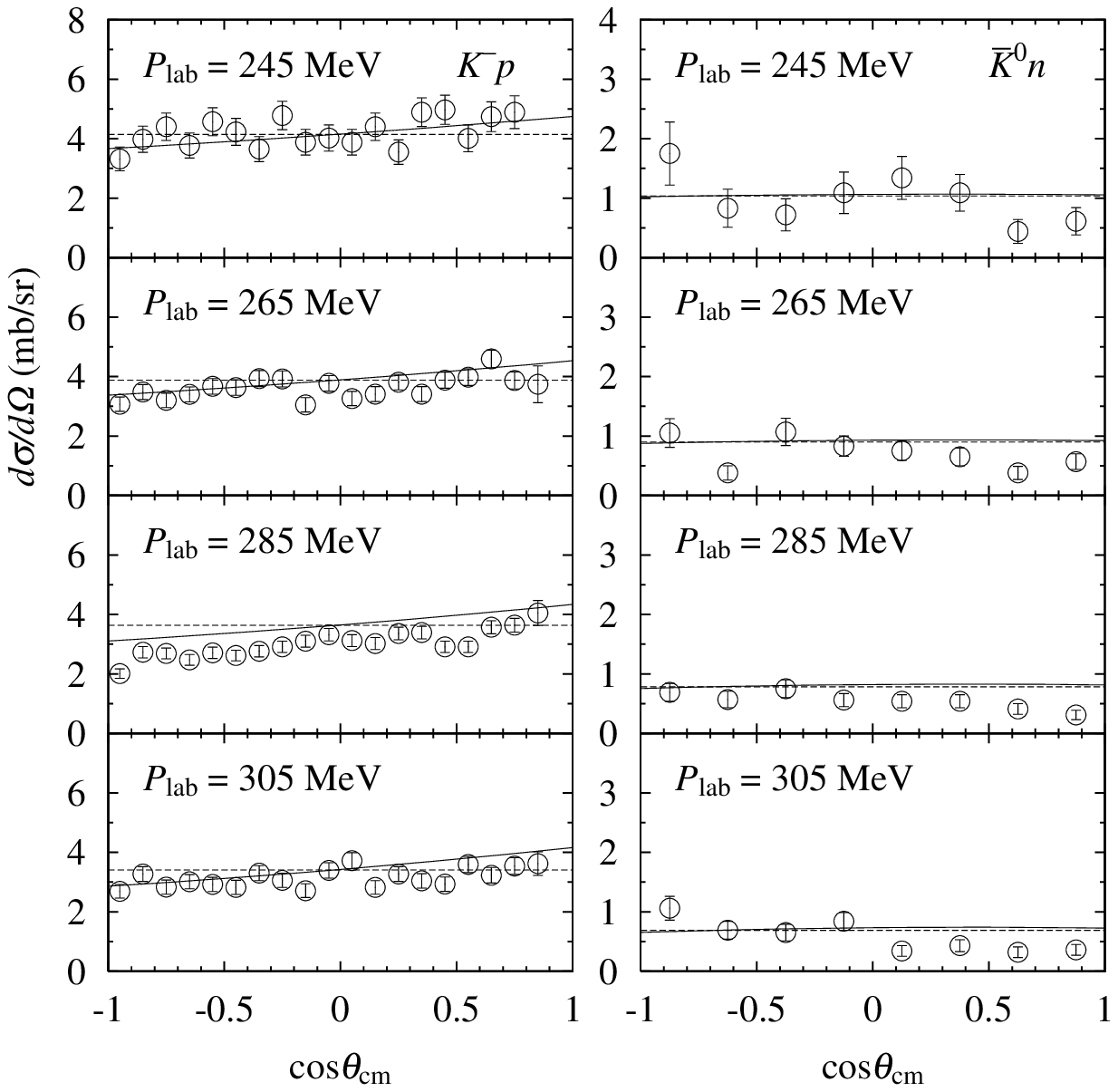}
   \end{center}
   \caption{
    Same as Fig.~\ref{fig:difforig} with the parameter set of
   Eq.~(\ref{subtsnew}).  
    \label{fig:diffnew}}  
\end{figure}

The effect of adding the \pwave\ is quite small in the cross sections,
justifying the success of the results obtained using the \swave\
alone. In order to better appreciate the effect of the \pwave, it is
better to look at the differential cross sections since there one is
sensitive to the interference of $s$- and {\pwave}s, which results in
larger effects than in the integrated cross section where just the
square of the \pwave\ amplitudes appears. We can see in
Fig.~\ref{fig:difforig} that the incorporation of the $p$-waves provides
the right slope in the differential cross sections, clearly indicating
that the amount of \pwave\ introduced is the correct one. The
agreement is not perfect for all laboratory momenta shown, but the
little strength missing or in excess is clearly due to the  dominant 
{\swave}. In order to emphasize this better we have taken advantage
of the fact that one can make a fine-tuning of the subtraction constants 
$a_{i}$ to improve the fit to the data. We have just changed the
parameters $a_{i}$ slightly to the values
\begin{eqnarray}
   a_{\bar K N} = -1.75, \hspace{0.5cm} a_{\pi \Sigma} = - 2.10, 
   \hspace{0.5cm} a_{\pi \Lambda} = -1.62 \nonumber \\
   a_{\eta \Lambda} = -2.56,  \hspace{0.5cm} a_{\eta \Sigma} =
   -1.54,  \hspace{0.5cm} a_{K \Xi} = -2.67 \label{subtsnew}
\end{eqnarray}
by means of which one obtains improved values for the low energy 
observables: 
\begin{equation}
    \gamma=2.36\ , \hspace{0.7cm} R_{c}=0.634\ , \hspace{0.7cm} 
    R_{n} = 0.178 \ .
\end{equation}
The values of the bare masses are now $\tilde M_{\Lambda} = 1069$ MeV,
$\tilde M_{\Sigma} = 1195$ MeV, $\tilde M_{\Sigma^{*}}=1413$ MeV. 
The results for the integrated cross sections with this new set of 
parameters are shown in 
Fig.~\ref{fig:totalnew}. The improvement is clearly appreciable in the
$K^{-}p \rightarrow K^{-}p$ and $K^- p \rightarrow \pi^0 \Lambda$ cross 
sections. The effect of the $p$-waves are more clearly shown
in Fig.~\ref{fig:diffnew}, where the 
differential cross sections for $K^{-} p \rightarrow K^{-} p$ and $K^{-} 
p \rightarrow \bar K^{0} n$ are now well reproduced. In fact, the {\pwave}s 
have 
barely changed from Fig.~\ref{fig:difforig} to Fig.~\ref{fig:diffnew}, 
but the slight improvement in the \swave\ brings the results in 
better agreement with experiment. It is interesting to mention 
that there has been no free parameter in the determination of the \pwave\ 
amplitude. The bare masses of the $\Lambda$, $\Sigma$, $\Sigma^{*}$ 
cannot be considered free parameters since they are determined by the 
physical masses of the baryons once the regularizing cut-off is chosen.

\section{Properties of the $\Sigma^{*}$(1385) resonance}%
\begin{figure}
   \begin{center}
   \epsfxsize=13cm
   \epsfbox{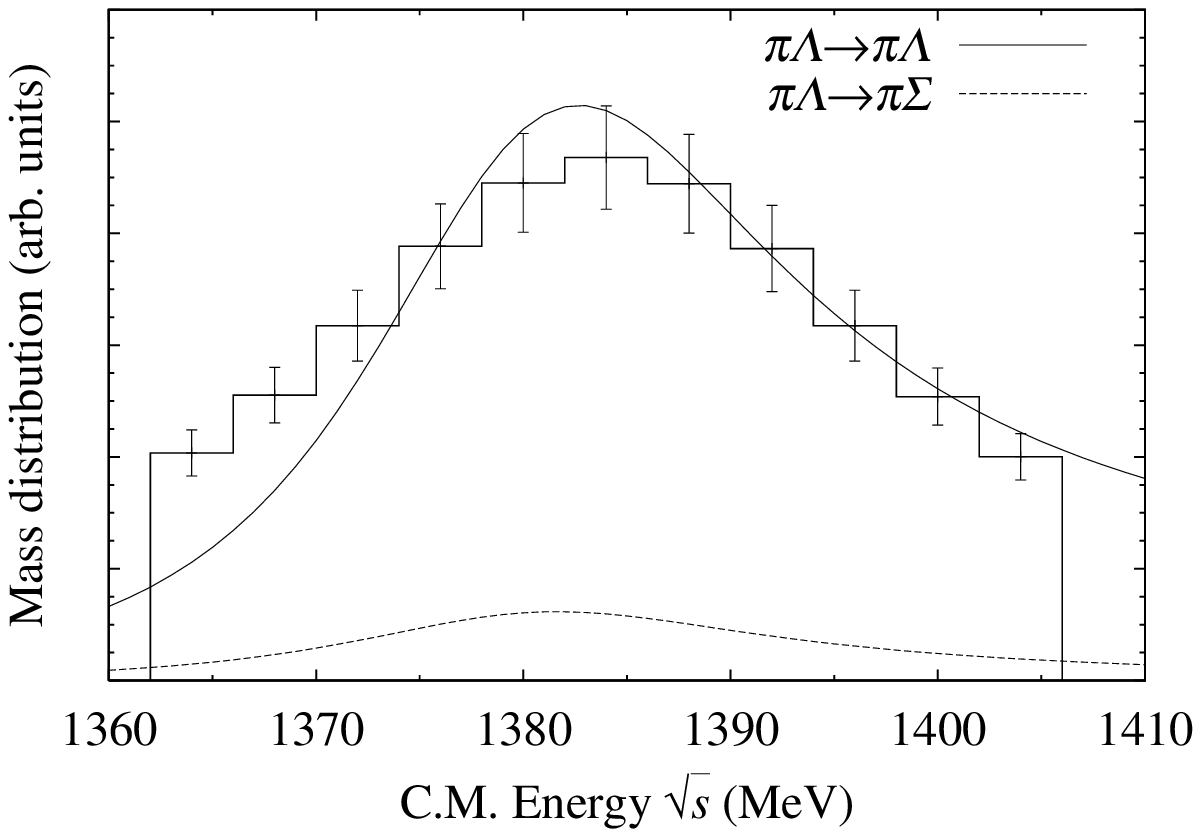}
   \end{center}
   \caption{$\Sigma^*(1385)$ mass distributions in arbitrary
   units. The solid lines denotes the mass distribution in the $\pi
   \Lambda \rightarrow \pi \Lambda$ reaction, while the dashed line
   shows the one for the $\pi \Sigma$ with $I=1$ in the final state. 
   The data are taken from \cite{Barreiro:1977uq}. \label{fig:massdist}}  
\end{figure}

We turn now to the results below threshold where the
$\Sigma^{*}$(1385) resonance appears. The results are seen in the
$\pi\Lambda$ or $\pi\Sigma$ 
mass distributions in reactions with 
$\pi\Lambda$ or $\pi\Sigma$ in the final state. The mass distribution 
of $\pi\Lambda$ is given by 
\begin{equation}
    {d \sigma \over d m } = C \left| t_{\pi\Lambda\rightarrow 
    \pi\Lambda}^{(I=1)} \right|^{2} p_{\rm CM}(\Lambda) \ ,
\label{Massdist}
\end{equation}
where the constant $C$ is related to the particular reaction generating 
the $\pi \Lambda$ state prior to final state interactions. Relatedly, the 
$\pi\Sigma$ mass distribution originating from the same primary mechanism 
will be given by Eq.~\ref{Massdist} by changing
$\left| t_{\pi\Lambda
\rightarrow \pi\Lambda}^{(I=1)} \right|^{2} p_{\rm CM}(\Lambda)$ by
$\left| t_{\pi\Lambda \rightarrow \pi\Sigma}^{(I=1)} \right|^{2}
p_{\rm CM}(\Sigma)$. The shape of the mass distribution is used
experimentally to obtain the position and width of the resonance, and
the ratio of partial widths of $\Sigma^{*} \rightarrow \pi\Lambda,
\pi\Sigma$ can be obtained by means of  
\begin{equation}
    {\Gamma_{\pi\Lambda} \over \Gamma_{\pi\Sigma}} = {
    \left| t_{\pi\Lambda \rightarrow \pi\Lambda}^{(I=1)} \right|^{2} 
    p_{\rm CM}(\Lambda) \over \left| t_{\pi\Lambda\rightarrow 
    \pi\Sigma}^{(I=1)} \right|^{2} p_{\rm CM}(\Sigma)}\ .
\end{equation}
In Fig.~\ref{fig:massdist} we can see the shape of the $\Sigma^{*}$ 
distribution, which a width of about $\Gamma_{\Sigma^{*}} \simeq 31$ 
MeV which compares favorably with the experimental value of 
$\Gamma_{\Sigma^{*}} \simeq 35 \pm 4$ MeV \cite{PDG}. The ratio of 
the partial decay widths obtained is 
\begin{equation}
    {\Gamma_{\pi\Lambda} \over \Gamma_{\pi\Sigma}} = 7.7 \ ,
\end{equation}
which compares well with the experimental value of $7.5\pm0.5$.

We have looked for poles in the complex plane in the \pwave\ 
amplitudes and have not found any, except for the $\Sigma^{*}$(1385) 
which is introduced as a genuine resonance in our approach, in the 
same way as the $\Lambda$ or $\Sigma$ baryons are included as basic
fields in the theory. This means that the strength of the lowest 
order \pwave\ amplitudes is too weak to generate dynamical 
resonances, contrary to what was found in \cite{cornelius} for the 
\swave s.

\section{Conclusions}
We have evaluated the \pwave\ amplitudes for meson-baryon scattering in
the strangeness $S=-1$ sector starting from the lowest order chiral
Lagrangian, unitarizing by means of the Bethe-Salpeter equation, or 
equivalently the $N/D$ method, and regularizing the loops with a cut-off 
or an equivalent
method using dimensional regularization.  The cut-off, or equivalently
the subtraction constants in the dimensional regularization procedure,
is fitted to the scattering observables at threshold.  In practice we take 
them from
earlier work \cite{cornelius} where only the \swave\ amplitudes were
studied in our approach.  Once this is done there is no extra freedom
for the \pwave\ amplitudes, which are completely determined in our
approach.  We perform some fine-tuning of the subtraction 
constants with respect to \cite{cornelius} in order to obtain better 
results for the $K^{-}p \rightarrow K^{-}p$ cross section, which 
however does not practically modify the \pwave\ amplitudes. 

The results which we obtain for the \pwave\ amplitudes in this 
approach are consistent with experimental data for the differential 
cross sections. The contribution to the total cross sections at low 
energies is very small but in the differential cross sections its 
effects are clearly visible producing a slope in $d \sigma / d \Omega$ 
as a function of $\cos \theta$ which is in good agreement with the data.

One of the most interesting features of the \pwave\ in the $S=-1$ 
sector is the presence of the $\Sigma^{*}$(1385) resonance below the 
$K^{-}p$ threshold. This resonance cannot be generated dynamically 
from the strength of the \pwave\ in lowest order of the chiral 
Lagrangians and hence is introduced as a basic field, like the 
$\Lambda$ or the $\Sigma$. It couples to the meson-baryon states with 
a strength obtained using SU(6) symmetry from the standard chiral 
Lagrangians involving pseudoscalar mesons and the octet of stable 
baryons. With these couplings and the unitarization procedure the 
$\Sigma^{*}$(1385) develops a width. In this sense, the total width of 
the $\Sigma^{*}$, as well as the branching ratios to $\pi\Lambda$ 
and $\pi\Sigma$, are predictions of the theory and they come out with 
values in agreement with experiment.

The approach followed here corroborates once more the potential of the 
chiral Lagrangians to describe the low energy interaction of mesons 
with baryons, provided a fair unitarization procedure is used to 
appropriately account for the multiple scattering of the many 
channels which couple to certain quantum numbers. In this particular 
case, the previous works in the $S=-1$ sector in \swave, together with 
the present one for \pwave, provide a good theoretical framework to study 
the meson-baryon dynamics at low energies. These works show that the basic 
dynamical information is contained in the chiral Lagrangian of lowest 
order, since by means of a proper unitarization procedure in coupled 
channels and one regularizing parameter of natural size for the loops,
one can describe quite well the low energy scattering data in the
different reactions with $S=-1$. 

\subsection*{Acknowledgments}

E. O. and D. J. wish to acknowledge the hospitality of the University
of Barcelona.  This work is also
partly supported by DGICYT contract numbers BFM2000-1326, PB98-1247,
by the EU TMR network Eurodaphne, contract no. ERBFMRX-CT98-0169, and
by the Generalitat de Catalunya project 2001SGR00064.


\end{document}